\documentclass[prb,twocolumn,showpacs,amsmath,amssymb]{revtex4-1}
\usepackage{graphicx}       
\usepackage{epsfig}
\usepackage{ulem}
\usepackage{dcolumn}        
\usepackage{color}
\newcommand{\beq}{\begin{equation}}
\newcommand{\eeq}{\end{equation}}
\newcommand{\kvec}{{\bf k}}
\newcommand{\qvec}{{\bf q}}   
\newcommand{\Qvec}{{\bf Q}}

\begin{document}
\title{Strange metal behaviour from charge density fluctuations in cuprates}

\author{G\"otz Seibold$^{1,*}$, Riccardo Arpaia$^{2,3}$, Ying Ying Peng$^{2, \dag}$, Roberto Fumagalli$^2$, Lucio Braicovich$^{2,4}$, 
Carlo Di Castro$^5$, Marco Grilli$^{5,6,*,\ddagger}$, Giacomo Claudio Ghiringhelli$^{2,7}$, and  Sergio Caprara$^{5,6,*,\ddagger}$}

\affiliation{$^1$ Institut f{\"u}r Physik, BTU Cottbus-Senftenberg - PBox 101344, D-03013 Cottbus, Germany} 

\affiliation{$^2$Dipartimento di Fisica, Politecnico di Milano, Piazza Leonardo da Vinci 32, I-20133 Milano, Italy}

\affiliation{$^3$Quantum Device Physics Laboratory, Department of Microtechnology and Nanoscience, 
Chalmers University of Technology, SE-41296 G{\"o}teborg, Sweden}

\affiliation{$^4$ESRF, The European Synchrotron, 71 Avenue des Martyrs, F-38043 Grenoble, France}

\affiliation{$^5$Dipartimento di Fisica, Universit\`a di 
Roma Sapienza, P.$^{le}$ Aldo Moro 5, I-00185 Roma, Italy}

\affiliation{$^6$CNR-ISC, via dei Taurini 19, I-00185 Roma, Italy}

\affiliation{$^7$CNR-SPIN, Dipartimento di Fisica, Politecnico di Milano, Piazza Leonardo da Vinci 32, 
I-20133 Milano, Italy}

\affiliation{$^{\dag}$Present address: International Center for Quantum Materials, School of Physics, Peking University, 
CN-100871 Beijing, China} 

\affiliation{$^*$Corresponding authors. E-mail: marco.grilli@roma1.infn.it \, sergio.caprara@roma1.infn.it \, seibold@b-tu.de}
\affiliation{$^\ddagger$ These authors jointly supervised this work}

\begin{abstract}{   {\bf Abstract~}
Besides the mechanism responsible for high critical temperature superconductivity, the grand unresolved issue 
of the cuprates is the occurrence of a strange metallic state above the so-called pseudogap temperature 
$T^*$. Even though such state has been successfully described within a phenomenological scheme, the so-called 
Marginal Fermi-Liquid theory, a microscopic explanation is still missing. However, recent resonant X-ray 
scattering experiments identified a new class of charge density fluctuations  characterized by low characteristic 
energies and short correlation lengths, which are related to the well-known charge density waves. These 
fluctuations are present over a wide region of the temperature-vs-doping phase diagram and extend well above 
$T^*$. Here we investigate the consequences of charge density fluctuations on the electron and transport properties 
and find that they can explain the strange metal  phenomenology. Therefore, charge density fluctuations are 
likely the long-sought microscopic mechanism underlying the peculiarities of the metallic state of cuprates.}
\end{abstract}

\maketitle

\noindent {\bf Introduction ~}
\par
\begin{figure*}
\includegraphics[angle=0,scale=0.35]{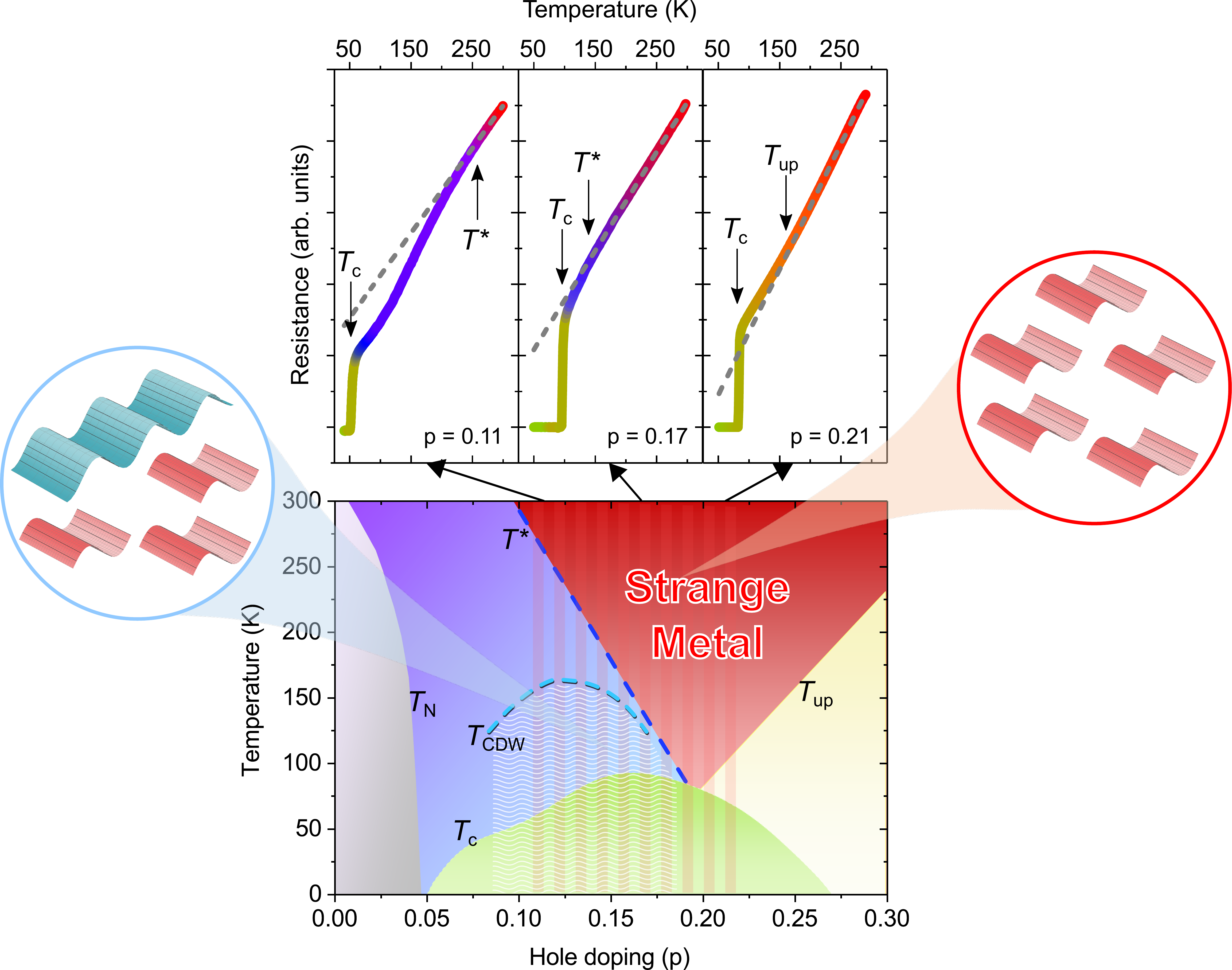}
\caption{{\bf Figure 1:} {{{{\bf Temperature-vs-doping phase diagram of the superconducting cuprates.} In the red region 
encompassed between the pseudogap temperature $T^*$ and the upturn temperature $T_{\mathrm{up}}$ of the resistance, 
above the superconducting critical temperature $T_{\mathrm{c}}$, in particular close to the optimally 
doped regime (e.g., at hole doping $p \approx 0.17$), these compounds display a strange metal  behaviour. This}}
is revealed in the experimental resistance $R$ data by the presence of a linear temperature dependence, displayed 
as a red thick solid line in the $R(T)$ curves above the phase diagram. In the underdoped regime (e.g., at 
$p \approx 0.11$), below $T^*$ (blue region) a downturn from the linear-in-$T$ resistance is 
observed, since additional mechanisms lead to deviations from the strange metal  regime. In the overdoped regime 
(e.g., at $p \approx 0.21$), below $T_{\mathrm{up}}$ (yellow region) the upturn from the linear-in-$T$ resistance  
is due to the setting in of the Fermi-liquid regime. Recent Resonant X-Ray Scattering experiments 
\cite{arpaia-2018} showed that also the charge order phenomenon is widespread in the phase diagram. In particular, 
short-ranged dynamical charge density fluctuations (sketched by red waves highlighted in the red circle, and observed 
in the striped area) populate the strange metal region, while in the underdoped region, below the onset temperature 
$T_{\mathrm{CDW}}$, they coexist with the usual longer-ranged charge density waves (sketched by blue waves in the 
blue circle, and observed in the wavy area). $T_\mathrm{N}$ is the N\'{e}el temperature. The data of the $R(T)$ curves 
are taken from Refs.\,\onlinecite{arpaia-2018, arpaia2018probing}.}}
\label{exp-LR}
\end{figure*}

\noindent
Among the different phases and orders populating the phase diagram of superconducting cuprates, the region where 
the strange metal occurs has a preeminent role for this class of compounds over a rather wide doping range 
pivoting around optimal doping (see Fig.\,\ref{exp-LR}). Experimentally, the most evident benchmark of this region 
is represented by the linear behaviour of the electrical resistivity $\rho(T)$ as a function of the temperature $T$, 
from above a doping-dependent pseudogap crossover temperature $T^*$ up to the highest attained temperatures. Such 
occurrence is less evident in the underdoped regime, where $T^*$ is almost as high as room temperature (e.g., at 
doping $p \approx 0.11$, see Fig.\,\ref{exp-LR}), while it dominates the transport properties of the metallic state in 
its entirety above optimal doping ($p \approx 0.17-0.20$, see Fig.\,\ref{exp-LR}), where $T^*$ decreases and 
eventually merges with the superconducting critical temperature $T_{\mathrm{c}}$. Beyond such occurrence, the 
main deviations from the paradigmatic behaviour dictated by the Landau Fermi-liquid theory of standard metals are 
the optical conductivity, following a non-Drude-like frequency dependence $\sigma(\omega)\sim 1/\omega$, and the 
Raman scattering intensity, starting linearly in frequency and then saturating into a flat electron continuum, 
as expressed by the dependence of the susceptibility of the scattering mediator, 
$\mathrm{Im}\,P(\omega) \sim \omega/ \mathrm{max}\,(T,|\omega|)$. It was shown long ago \cite{MFL} 
that the phenomenological assumption of this form for $\mathrm{Im}\,P(\omega)$ accounts for the above anomalous 
properties. In particular, the related low-energy excitations, mediating a momentum-independent electron-electron 
effective interaction, give rise to a linear dependence of the imaginary part of the electron self-energy both in 
frequency and temperature
\beq
\mathrm{Im}\,\Sigma (\kvec,\omega) \sim \mathrm{max}\,(T,|\omega|).
\label{imsigma}
\eeq
Although there are theories that do not rely on a specific mediator \cite{kastrinakis}, 
a huge effort has been devoted along the years to identify the excitations mediating this scattering, mostly based 
on the idea of proximity to some form of order: circulating currents \cite{varma-2007}, spin 
\cite{abanov,norman-chubukov}, charge order \cite{CDG-1995,reviewQCP1,kivelson_review,caprara-2016,cdfg-2002}, or 
the phenomenological coupling to incoherent fermions \cite{sachdev-PRX}. 

A step forward in the identification of low-energy excitations that might be responsible for the strange 
metal behaviour was recently taken by means of resonant X-ray scattering (RXS), performed on 
Nd$_{1+x}$Ba$_{2-x}$Cu$_3$O$_{7-\delta}$ (NBCO) and YBa$_2$Cu$_3$O$_{7-\delta}$ (YBCO) thin films \cite{arpaia-2018}. 
After the first experimental evidence, these excitations have been demonstrated to be a common feature of different 
families of cuprates, namely HgBa$_2$CuO$_{4+\delta}$ \cite{greven-2019}, La$_{2-x}$Sr$_x$CuO$_4$ 
\cite{dean1-2020,dean2-2020,wen-2019}, La$_{2-x}$Ba$_{x}$CuO$_4$ \cite{miao17,miao19}, and 
La$_{1.675}$Eu$_{0.2}$Sr$_{0.125}$CuO$_4$ \cite{chang-2020}, thereby indicating that these excitations
may well provide a generic scattering mechanism in all cuprates.

In the following we will focus on NBCO or YBCO investigated in the precursor experiment. These experiments not only 
confirmed the occurrence of incommensurate charge density waves (CDWs), correlated over several lattice spacings, in 
the underdoped region and below $T^*$ \cite{allRIXS,achkar2012distinct,tabis2014charge,comin2014charge,
blanco2014resonant,keimer,gerber2015three,comin2016resonant,peng2018re}, but, quite remarkably, 
also identified a much larger amount of very short-ranged ($\approx 3$ lattice spacings) dynamical charge density 
fluctuations (CDFs, see Fig.\,\ref{exp-LR}), with a characteristic energy scale $\omega_0 \approx 10-15$\,meV. 
These CDFs are peaked at a wave vector, along the (1,0) and (0,1) directions, which is very close to that of the 
intermediate-range CDWs\cite{arpaia-2018}, arising below a given temperature $T_\mathrm{CDW}(p)$ for each measured 
doping $p$. We also notice that, when the temperature is raised towards $T_\mathrm{CDW}(p)$, the CDWs 
correlation length decreases down to values close to those of the CDFs. These facts suggest that the two charge 
fluctuations have a common origin. One possibility is that 
they develop differently in different regions, with CDFs remaining non critical, whereas CDWs evolve towards order. 
This is also supported by the possibility that the narrow peak (NP) of the RXS 
response function, customarily associated to the CDWs, arises at the expense of the broad peak (BP) due to CDFs. 
However, differently from CDWs, CDFs are quite robust both in temperature (they survive essentially unaltered up to 
the highest explored temperatures, $T\approx 270$\,K) and doping. These excitations are at low energy ($\approx 15$\,meV 
in an optimally doped sample with $T_{\mathrm c}=90$\,K) and so short ranged that in reciprocal space they produce the 
BP observed in the RXS scans. CDFs not only provide a strong scattering channel for the electrons, but also overcome 
the difficulty of the CDWs, which, being so peaked, give rise to anisotropic scattering dominated by the hot spots on 
the Fermi surface. CDFs, instead, being so broad, affect all states on the Fermi surface nearly equally, resulting in 
an essentially isotropic scattering rate. This isotropy is a distinguished feature of the strange metal state and we 
show below that it can account for the peculiar behavior of the electronic spectra
and for the linear-in-$T$ resistivity.
\vspace{1 truecm}

\noindent {\bf Results}

\noindent {\bf \small Strange Metal behaviour of the electron self-energy.~}
\begin{figure*}
\includegraphics[angle=0,scale=0.45]{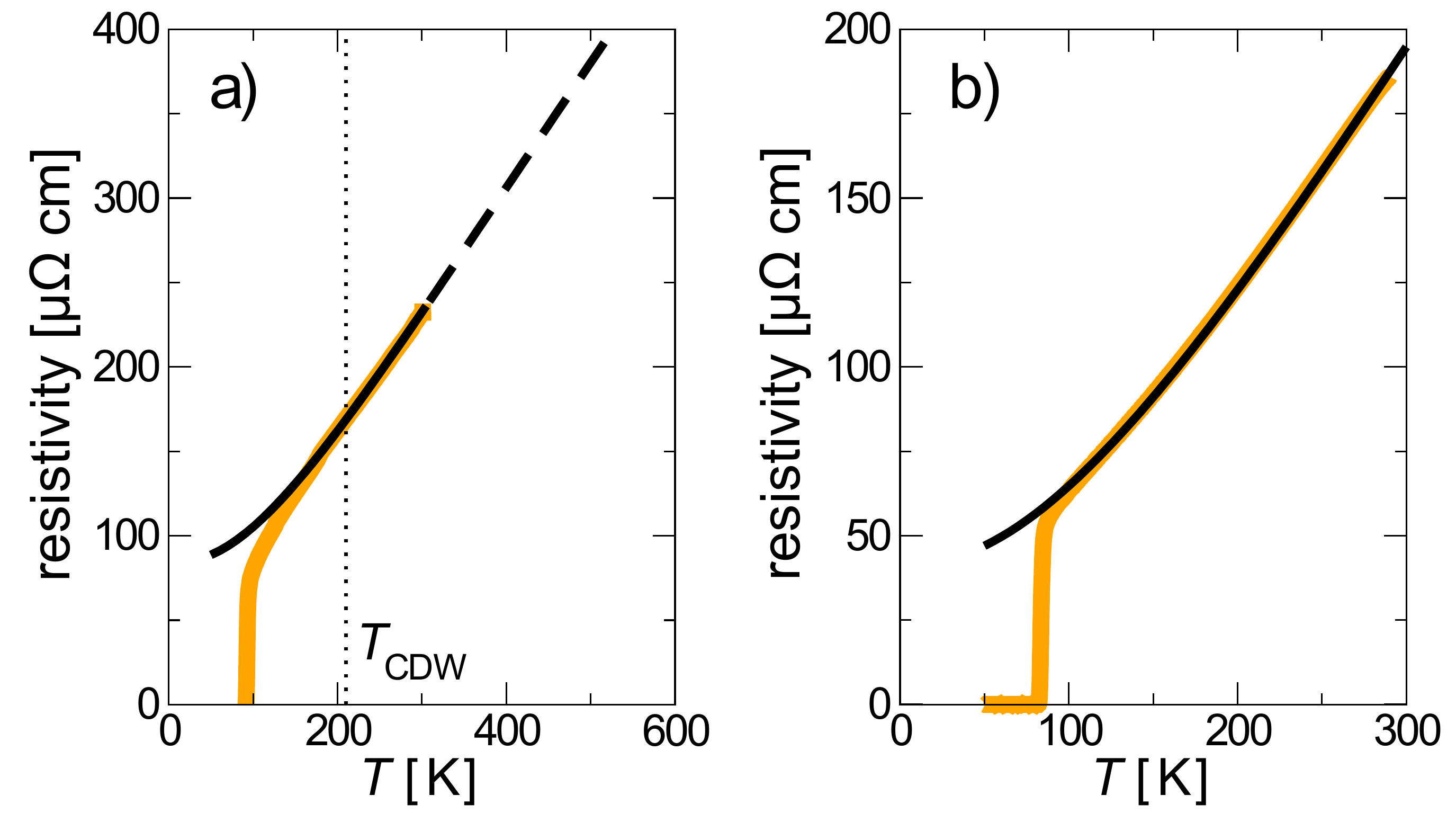}
\caption{{\bf Figure 3:} {{\bf Linear-in-$T$ resistivity.} {\bf (a)} Experimental resistivity for an optimally doped 
($T_{\mathrm c}=90$\,K) Nd$_{1+x}$Ba$_{2-x}$Cu$_3$O$_{7-\delta}$  sample (yellow thick curve) 
compared to the theoretical result as obtained from the charge density fluctuations (CDFs) only (black solid line). 
The dashed part demonstrates the continuation of linear-in-$T$ behaviour up to temperatures $>500$\,K. The scattering 
rate includes an elastic scattering $\Gamma_0$ due to quenched impurities, 
$\Gamma(\phi)=\Gamma_0 + \mathrm{Im}\,\Sigma (\phi,T,\omega=0)$. Here, { {$\Gamma_0=52$\,meV}}, and the coupling 
$g=0.166$\,eV between quasiparticles and CDFs is the same as for the self-energy of Fig.\,\ref{fig-self-en}. {\bf (b)} 
Same as (a) for an overdoped YBa$_2$Cu$_3$O$_{7-\delta}$ sample ($T_{\mathrm c}=83$\,K). Here, 
$\Gamma_0=25.5$\,meV, $g=0.179$\,eV.}}
\label{rhoT}
\end{figure*}
Fig.\,\ref{fig-self-en}(a) shows a qualitative explanation of the inherent isotropy of the scattering by CDFs.
RXS experiments directly access the frequency and momentum-dependent charge susceptibility (see Methods)
and find the above mentioned BP at a well defined incommensurate wave vector $\Qvec_c$, but the large width
of this peak means that a wealth of low-energy CDFs are present over a broad range of momenta. Therefore, an 
electron quasiparticle on a branch of the Fermi surface can always find a CDF that scatters it onto another 
region of the Fermi surface [see Fig.\,\ref{fig-self-en}(a)]. Thus the whole Fermi surface is 
hot in the sense that no regions exist over the Fermi surface that can avoid this scattering. This is 
visualized in Fig.\,\ref{fig-self-en}(a), where the overlap of the Fermi surface with its translated and broadened 
replicas (due to the scattered quasiparticles) is almost uniform, and no particular nesting condition is needed.
Quite remarkably, the CDF-mediated scattering stays isotropic even in an energy window of 
$\approx 20$\,meV around the Fermi surface (see supplementary note 1 and 
supplementary figure 2).

On the contrary, since CDWs are quite peaked, only a few of them around $\Qvec_c$ are available to scatter 
quasiparticles at low energy: Only quasiparticles at the hot spot are then significantly scattered by CDWs 
[see Fig.\,\ref{fig-self-en}(a)]. In a quantitative way, this is shown in Fig.\,\ref{fig-self-en}(b), where the 
actual scattering rate along the Fermi surface has been separately computed for CDFs (solid red line) and CDWs 
(dashed blue line) with parameters suitable to describe a slightly underdoped NBCO sample ($p\approx 0.15$), where 
CDF and CDW coexist (see supplementary note 2 and supplementary figure 5). This feature makes these CDFs an 
appealing candidate to mediate the isotropic scattering required by the original marginal Fermi-liquid theory. 
We therefore test this expectation by explicitly calculating how the CDFs dress the electron quasiparticles 
modifying their spectrum. In many-body theory, this effect is customarily described by the electron self-energy. 
In particular, the imaginary part of the electron self-energy, $\mathrm{Im}\,\Sigma$, provides the broadening of 
the electron dispersion as measured, e.g., in angle-resolved photoemission experiments. We adopt the following 
strategy: a) we extract from the experimental inelastic RXS spectra the information on the dynamics of the CDFs 
(see supplementary note 2) evaluated within the linear response theory; b) we borrow from photoemission experiments 
the electron dispersion in the form of a tight-binding band structure \cite{meevasana}; 
c) we calculate the electron self-energy resulting from the coupling between CDFs and the electron quasiparticles, 
as discussed in supplementary note 1 and represented as a diagram in supplementary figure 1.

With the extracted parameters, using the coupling between quasiparticles and CDFs obtained from the 
resistivity fit (see below) and taking the frequency derivative of the real part of the self-energy, 
we also calculated the dimensionless coupling $\lambda$ at $T\approx T^*$ finding $\lambda\approx 0.35-0.5$ 
(see supplementary note 1).

Of course, this perturbative approach, although supported by the 
low-moderate value of $\lambda$, is based on the Fermi liquid as a starting point in the overdoped region. 
Its applicability can be safely extended to lower doping at high temperatures, in the metallic state and 
above $T^*$, where the phenomenology is only marginally different from that of a Fermi Liquid. 

The result of our calculation for an optimally doped NBCO sample with $T_\mathrm{c}=90$\,K is reported in
Figs.\,\ref{fig-self-en}(c,d). After an initial quadratic behaviour, the scale of which is set by
the energy scale $\omega_0$ of the CDFs \cite{caprara99} (see supplementary note 1), $\mathrm{Im}\,\Sigma$ 
displays an extended linear frequency dependence up to $0.10-0.15$\,eV (comparable to the one reported in 
the photoemission experiments of Refs.\,\onlinecite{valla-1999,bok-2010}). The overall value of this self-energy 
is comparable to, but it always stays smaller than, the Fermi energy scale of order $0.3-0.4$\,eV. This is an 
intrinsic manifestation of a strange metal state, where the width of the quasiparticle peak must be of the same 
order of its typical energy. At low frequencies $\mathrm{Im}\,\Sigma$ saturates at a constant value that increases 
linearly with increasing $T$. This is precisely the behaviour expected from the strange metal expression of 
Eq.\,(\ref{imsigma}). This self-energy is reported along a specific (1,1) direction, but it is crucial to recognize 
that it is also highly isotropic in momentum space. Fig.\,\ref{fig-self-en}(b) indeed reports the scattering 
rate (i.e., the imaginary part of the self-energy at zero frequency) 
$\Gamma(\phi)\equiv\Gamma_0 + \Gamma_{\Sigma}(\phi)$. An isotropic scattering rate $\Gamma_0$ representing the 
effect of quenched impurities has also been included. Our results in Fig.\,\ref{fig-self-en}(c), not only share 
with the data of Ref.\,\onlinecite{valla-1999} a similar form, but also display a scaling behaviour, as reported 
in Fig.\,\ref{fig-self-en}(d). As mentioned in Ref.\,\onlinecite{MFL}, the isotropic linear-in-frequency 
self-energy behaviour, stemming from CDFs, is sufficient to produce a strange metal behaviour in physical 
quantities like optical conductivity and Raman scattering.

Below $T_{\mathrm{CDW}} =150$\,K, an additional scattering due to the CDWs is present. 
This additional scattering has a significant anisotropic component, which is confined in a small region 
of momentum space, as shown by the dashed blue curve of Fig.\,\ref{fig-self-en}(b). This anisotropic character 
eventually leads to the departure from the strange metal behaviour \cite{hlubina} below temperatures comparable 
with $T^*$.

\vspace{1 truecm}
\noindent
{\bf \small CDFs produce linear resistivity.~}
\begin{figure*}
\includegraphics[angle=0,scale=0.55]{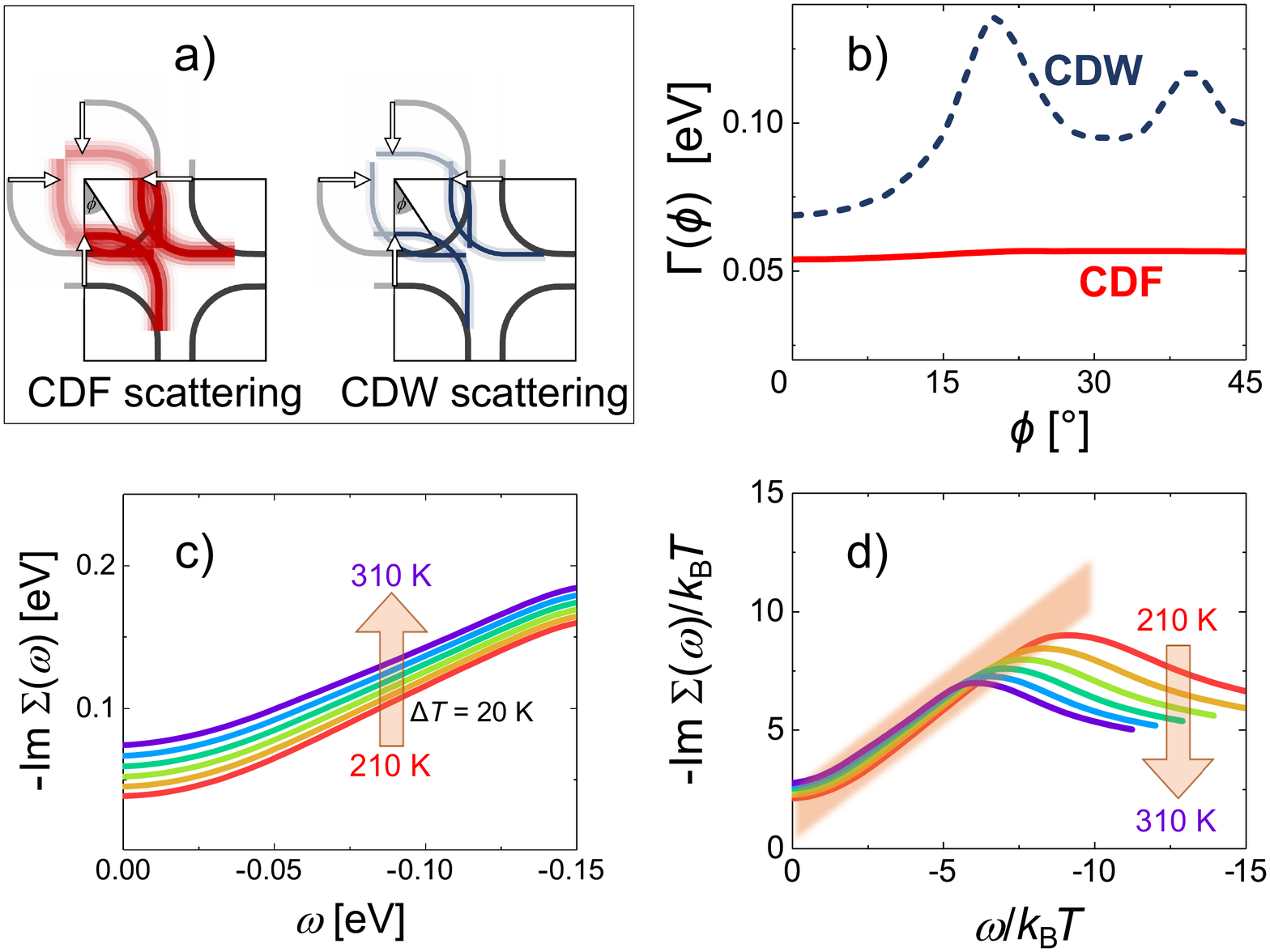}
\caption{{\bf Figure 2:} {{\bf Strange metal self-energy.} {\bf (a)} Sketch of the charge density fluctuation (CDF) and 
charge density wave (CDW) mediated quasiparticle scattering on the Fermi surfaces. Points on the Fermi surface are 
identified by the angle $\phi$. Owing to the broadness of CDFs in momentum space, all the states along the Fermi 
surface (thick black line) can be scattered by low-energy CDFs over other portions of the Fermi surface, and no 
particular nesting condition is needed. The involvement of only one branch of the Fermi surface in the Brillouin 
zone is displayed for clarity: The scattered portions of the Fermi surface (broad reddish areas) essentially cover 
the whole branch. Therefore the whole Fermi surface is affected in a nearly isotropic way. On the contrary, the CDWs 
are peaked in momentum space and scatter the Fermi surface states in rather restricted regions of other Fermi surface 
branches (hot spots). These occur where the bluish lines cross the thick black line. {\bf (b)} Scattering rate 
[i.e., the imaginary part of the self-energy at zero frequency $\Gamma(\phi)=-\mathrm{Im}\,\Sigma (\phi,T,\omega=0)$] 
at a given temperature $T=80$\,K, as a function of the position on the Fermi surface, as identified by the angle 
$\phi$ defined in panel (a). The nearly isotropic red line corresponds to the case when all the scattering would be 
due to CDFs, while the blue dashed line represents the scattering due to CDWs only. {\bf (c)} Imaginary part of 
the electron self-energy as a function of the (negative) electron binding energy, at different temperatures above 
$T_{\mathrm{CDW}}$, below which the CDWs emerge to produce the narrow peak in resonant X-ray scattering. The coupling 
between fermion quasiparticles and CDFs is $g=0.166$\,eV. {\bf (d)} Same as (c), but with both frequency and self-energy 
axes rescaled by the temperature ($k_{\mathrm B}$ is the Boltzmann constant), to highlight the approximate scaling 
behaviour at low frequency}
}
\label{fig-self-en}
\end{figure*}
\noindent Once the dynamics of the CDFs is identified by exploiting RXS experiments, one can investigate their 
effects on transport properties. The calculation of the electron resistivity is carried out within a standard 
Boltzmann-equation approach along the lines of Ref.\,\onlinecite{hussey} (see supplementary note 3). An analogous 
calculation within the Kubo formalism gives very similar results (see supplementary note 4 and supplementary 
figure 7). From the electron self-energy we obtain the zero frequency quasiparticle scattering rate along the 
Fermi surface $\Gamma(\phi)$ defined above, and we use $\Gamma_0$ as a fitting parameter, obtaining values 
($\approx 30-60$\,meV) that are reasonable for impurity scattering. We also use the anisotropic Fermi wave 
vector along the Fermi surface, as obtained from the same band structure in tight-binding approximation 
\cite{meevasana} used for the self-energy calculation. Fig.\,\ref{rhoT}(a) displays the comparison between 
the $\rho(T)$ curve of the optimally doped NBCO film ($T_{\mathrm{c}}=90$\,K), studied in Ref.\,\onlinecite{arpaia-2018} 
(yellow line) and the theoretical results (black line). At high temperatures, the famous linear-in-$T$ behaviour 
of the resistivity is found and the data are quantitatively matched. This behaviour stems from the very isotropic 
scattering rate produced by the CDFs [red solid line in Fig.\,\ref{fig-self-en}(b)], which, for this sample and in 
this temperature range, are  the only observed charge excitations. At lower temperatures, below $T^*$, a discrepancy 
emerges between the theoretical expectation and the experimental evidence, since the expected saturation, due to 
the onset of a Fermi-liquid regime and to (isotropic) impurity scattering $\Gamma_0$, is experimentally replaced 
by a downturn of the resistivity. Such discrepancy occurs gradually in $T$ when, entering the pseudogap state, 
the pseudogap itself and other intertwined incipient orders (CDWs, Cooper pairing,...) play their role.
These effects, which are outside our present scope, obviously lead to deviations from our theory, which only 
considers the effect of CDFs. On the other hand, in the overdoped YBCO sample 
($T_{\mathrm{c}}=83$\,K), the pseudogap and the intertwined orders are absent, while the CDFs are the only surviving 
charge excitations, even down to $T_\mathrm{c}$ \cite{arpaia-2018}. Here, our theoretical resistivity, related to the 
scattering rate produced by CDFs, matches very well the experimental data, in the whole range from room temperature 
almost down to $T_{\mathrm{c}}$ [see Fig.\,\ref{rhoT}(b)]. In particular, the agreement is rather good even at the 
lowest temperatures above $T_\mathrm{c}$.  The data display an upward saturation due to the onset of a 
Fermi-liquid regime that is well described by our calculation: At temperatures lower than the characteristic energy 
of CDFs their scattering effect is suppressed and the strange-metal behaviour ceases. We find remarkable that our theory 
not only describes the linear-in-$T$ regime, but also captures the temperature scale of upward deviation from it, 
without additional adjustments.

\vspace{1 truecm}
\noindent
{\bf Discussion and conclusions}

\noindent
The above results clearly show that the main features for the CDFs to account for the strange metal behaviour 
are a) a short coherence length of $1-2$ wavelengths to scatter the low energy electrons in a nearly isotropic way 
and b) a rather low energy ($\omega_0\approx 10-15$\,meV) to produce a linear scattering rate down to $100-120$\,K. 
We emphasize here that $\omega_0$ is only a characteristic minimal scale of CDFs, but these are broad overdamped
excitations from $\omega=0$ (due to damping) up to about 0.1\,eV, because they have a dispersion 
${\sim\overline{\nu}} (\qvec-\Qvec_c)^2$ with a stiffness energy scale 
${\overline{\nu}}\approx 1.0-1.5$\,eV(r.l.u.)$^{-2}$ [see Eq.\,(\ref{fluctuator}), the discussion in
supplementary note 2, and supplementary figures 5 and 6].  Moreover, our approach (extract information about 
CDFs from RXS experiments, and determine their effect on electron spectra and transport), not only captures the 
high-temperature linear behaviour of resistivity, but also the deviation from it in the overdoped case, where no 
other perturbing mechanisms, like CDWs, pairing, spin fluctuations, pseudogap, are present.

The question may also arises whether CDFs can also account for the so-called Planckian behaviour \cite{planckian}: 
at some specific doping, when a strong magnetic field (several tens of Teslas) destroys superconductivity, the 
linear-in-$T$ resistivity extends down to low temperatures of a few K. In order for our theory to account also 
for this behaviour, we should find CDFs with a lower characteristic energy of order $0.5-1.0$\,meV, while 
maintaining the correlation length short, to keep the scattering isotropic. Unfortunately at the moment no RXS 
experiments in the presence of such large magnetic fields are viable and we therefore cannot test these expectations. 
Nevertheless, we feel that it is not accidental that our theory accounts so well of the experiments done so far in 
the absence of magnetic field which show linearity up to very high temperature, well above $T^*$, so far from the 
quantum region. No wonder if by lowering the temperature at special values of doping, other effects may come in 
to modify our parameters values.

One interesting question is why CDFs, even in the absence of the specific Planckian conditions
have rather low characteristic energies $\approx 10-15$\,meV. In this regard, we notice 
that CDFs and CDWs have nearly the same characteristic wave vectors, indicating a close relationship. Since CDWs 
have a nearly critical character (that was theoretically predicted long ago \cite{CDG-1995,andergassen}), it 
is likely that CDFs are aborted CDWs, that for several possible reasons (competition with superconductivity, 
low dimensionality, disorder, charge density inhomogeneity, ...) do not succeed in establishing longer-range 
correlations. Still, this tight affinity with CDWs, which are nearly critical and therefore at very low energy, 
implies that CDFs also may have a broad dynamical range extending down to a rather low energy scale $\omega_0$. 
In this scenario, where CDWs and CDFs coexist in the system, one and the same theoretical scheme accounts for both 
excitations.

In conclusion, although some issues are still open, like the effects of magnetic field on CDFs to possibly account for
Planckian transport, or the  origin of the pseudogap features in transport, we were able to show that CDFs account for 
the anomalous metallic state of cuprates above $T^*$. Indeed, once the dynamics of the CDFs is extracted from RXS 
experiments, we can well explain, with the same parameter set, both the strange metal behaviour of the electron 
self-energy (therefore all the related anomalous spectral properties observed, e.g., in optical conductivity and Raman 
spectroscopy, are also explained) and the famous linear-in-$T$ resistivity in the metallic state of high-temperature 
superconducting cuprates. We thus believe that our results provide a very sound step forward in the long-sought 
explanation of the violation of the normal Fermi-liquid behaviour in cuprates.

\vspace{1 truecm}
\par\noindent
{\bf {Methods}}

\noindent
{\bf \small Fitting procedure to extract the CDW and CDF dynamics.~}
The CDW and CDF contributions to the RXS spectra are captured by a density
response-function diagram as reported in supplementary figure 1(a). In this framework, we carry out a 
twofold task: on the one hand, we show that dynamical CDFs and nearly critical CDWs
account both for the RXS high-resolution, frequency dependent, spectra, and for the quasi-elastic
momentum-dependent spectra. On the other hand, from the fitting of these experimental quantities,
we extract the dynamical structure of these excitations needed to calculate the physical quantities
discussed above.

According to this scheme, the CDW or CDF contribution to the low-energy RXS spectra is 
\beq
I(\qvec,\omega)=A\,\mathrm{Im}\,D(\qvec,\omega)\,b(\omega)
\label{teo-HR}
\eeq
where $b(\omega)\equiv[\mathrm e^{\omega/{k_BT}}-1]^{-1}$ is the Bose distribution 
ruling the thermal excitation of CDFs and CDWs, and $A$ is 
a constant effectively representing the intricate photon-conduction electron scattering processes 
\cite{comin2016resonant,reviewRIXS}.
In Eq.\,(\ref{teo-HR}), $\mathrm{Im}\,D(\qvec,\omega)$ is the imaginary (i.e., absorptive) part of the (retarded) 
dynamical density fluctuation propagator, which can describe either CDWs or CDFs. For both we adopt the standard 
Ginzburg-Landau form of the dynamical density fluctuation propagator, typical of overdamped quantum critical 
Gaussian fluctuations \cite{CDG-1995,reviewQCP1,andergassen},
\beq
D(\qvec,\omega)\equiv\left[\omega_0+\nu(\qvec)-\mathrm i\omega -\frac{\omega^2}{\overline\Omega} \right]^{-1},
\label{fluctuator}
\eeq
where $\omega_0=\bar\nu\, \xi^{-2}$ is the characteristic energy of the fluctuations, 
$\nu(\qvec)\approx \bar\nu\, |\qvec -\Qvec_\mathrm{c}|^2$, $\bar\nu$ determines the dispersion of the density 
fluctuations, $\Qvec_\mathrm{c}\approx (0.3,0),(0,0.3)$ is the characteristic critical wave vector (we work with 
dimensionless wave vectors, measured in reciprocal lattice units, r.l.u.) and $\overline\Omega$ is a frequency 
cutoff. This form of the charge collective mode propagator is typical of metallic systems where the collective 
modes have a marked overdamped character at low energy, where they can decay into particle-hole pairs (Landau 
damping). At larger energies, above $\overline\Omega$, they acquire a more propagating character. In both 
regimes, however, the maximum of their spectral weight is dispersive with a definite relation between $\omega$ 
and momentum, as it should be for well-defined collective modes. This is valid for both CDFs and CDWs, although 
the coherence length of the formers is weakly varying in doping and temperature and is generically very short (of 
the order of the wavelength itself). The sharper CDWs have a nearly critical character, with a marked temperature 
dependence of the square correlation length, $\xi^2_{\mathrm{NP}}(T)$. In particular, if these fluctuations had a standard 
quantum critical character around optimal doping \cite{CDG-1995,reviewQCP1,andergassen,CDSG-2017}, one would expect 
$\xi^2_{\mathrm{NP}}(T)\sim 1/T$. The CDFs have a similar $\Qvec_\mathrm{c}$, the main difference being in the behaviour of 
the correlation length, that, according to RXS experiments, increases significantly with decreasing the temperature 
and reaches up to $8-10$ lattice spacings for the nearly critical CDWs, while the CDFs have correlation length in the 
range $2-3$ lattice spacings, independently of the temperature. 

Although high-resolution spectra provide a wealth of information, they are experimentally very demanding,
so that RXS data are more often available in the form of quasi-elastic spectra corresponding to the 
frequency integration of the inelastic spectra, Eq.\,(\ref{teo-HR}),
\beq
I(\qvec)= \int_{-\infty}^{+\infty} 
\frac{A\,\omega}{\left( \omega_0+\nu (\qvec)  
-\frac{\omega^2}{\overline\Omega}\right)^2  +\omega^2} \,b(\omega) \,\mathrm d\omega  \label{teo-LR1} 
\eeq
Our first goal is to extract from the experiments all 
the parameters entering the CDW and CDF correlators, $\omega_0,\bar\nu,\Qvec_\mathrm{c}$ and $\overline\Omega$.

Since high-resolution and quasi-elastic spectra provide different complementary information, we adopted a 
bootstrap strategy in which we first estimated the dynamical scale $\omega_0$ from high-resolution at the 
largest temperatures, where the NP due to CDWs is absent and all collective charge excitations are CDFs. 
Then, we used this information to fit the quasi-elastic peaks to extract the relative weight (intensity) of 
the narrow and broad contributions. Once this information is obtained, we go back to high resolution spectra, 
since we now know the relative weight of the CDFs and CDWs contribution at all temperatures.

More specifically, the quasi-elastic peak has a composite character and, once the (essentially linear) background 
measured along the $(1,1)$ direction is subtracted (see, e.g., Fig.\,2 A-D in Ref.\,\onlinecite{arpaia-2018}), 
the peak may be decomposed into two approximately Lorentzian curves, corresponding to a narrow, strongly temperature 
dependent, peak due to the standard nearly critical CDWs arising below $T\approx 200$\,K and to a BP due to the 
CDFs. 
This is the main outcome of the RXS experiments reported in Ref.\,\onlinecite{arpaia-2018}. We thus fitted each 
of the two peaks {{with  equation (\ref{teo-LR1}).}} From the fits, one can extract the overall intensity 
parameter $A$ and the ratio $\omega_0/\bar\nu=\xi^{-2}$. Since only this ratio determines the width of the 
quasi-elastic spectra, we need a separate measure to disentangle $\omega_0$ and $\bar\nu$, so we used the 
high-resolution information on $\omega_0$ for the BP at $T=150$\,K and $T=250$\,K to extract 
$\bar\nu_{\mathrm{BP}}\approx 1400$\,meV(r.l.u.)$^{-2}$ at these temperatures. The same procedure cannot be 
adopted for the narrow CDWs peaks, which always appear on top of (and are hardly unambiguously separated from) 
the broad CDFs contribution. Nevertheless, to obtain a rough estimate, we investigated the high-resolution spectra 
at low temperature (see supplementary note 2), where the maximum intensity should mostly involve the NP to extract 
the characteristic energy of the quasi-critical CDWs obtaining, as expected, much lower values 
$\omega_0^{\mathrm{NP}}\approx 1-3$\,meV (although these low values are less reliable, due to the relatively low 
resolution of the frequency-dependent spectra). These estimates allow to extract values of 
$\bar\nu_{\mathrm{NP}}\approx 800$\,meV(r.l.u.)$^{-2}$ for the CDWs, comparable with those of the CDFs, suggesting 
a common electronic origin of the two types of charge fluctuations. To reduce the fitting parameters to a minimum, 
although subleading temperature dependencies of the high-energy parameters $\bar\nu$ and $\overline\Omega$ over a 
broad temperature range can be expected, we kept those parameters constant. We also assumed a constant $\omega_0$ 
for the CDFs, to highlight the non-critical nature of these fluctuations.

\vspace{1 truecm}
\noindent
{\bf {Data availability}}

\noindent
The  experimental resistivity and RXS data (see Fig.\,3 of the main text and supplementary figures 4-7) have already been 
published in Ref.\,\onlinecite{arpaia-2018} and are therefore available in the related data repository \cite{data_available}. 
They are also available from one of the corresponding authors [M.G.] on reasonable request. The datasets (resistivity 
curves, fitted RXS spectra, and electron self-energy) generated during the current study
are available from one of the corresponding authors [M.G.] on reasonable request.

\vspace{0.5 truecm}
\noindent
{\bf {Code availability}}

\noindent
The theoretical analysis was  carried out with FORTRAN codes to implement various required numerical integrations
[Eq.\,(\ref{teo-LR1}) in Methods to fit the RXS data, supplementary equation (1)
for the self-energy, in the supplementary note 1, and supplementary equation (6) for the resistivity, in supplementary 
note 3]. Although the same task could easily by performed with Mathematica or other standard softwares, the  
FORTRAN codes we used are available from one of the corresponding authors [M.G.] on reasonable request.


\vspace{1.5 truecm}

\noindent
{\bf {Acknowledgments}}
\par\noindent
We thank  C. Castellani, S. Kivelson, M. Le Tacon, M. Moretti Sala and T. P. Devereaux for  stimulating discussions.
We acknowledge financial support from the University of Rome Sapienza, through the projects Ateneo 2017 (Grant No. 
RM11715C642E8370), Ateneo 2018 (Grant No. RM11816431DBA5AF), Ateneo 2019 (Grant No. RM11916B56802AFE), from the Italian 
Ministero dell'Universit\`a e della Ricerca, through the Project No. PRIN 2017Z8TS5B, and from the Fondazione CARIPLO 
and Regione Lombardia, through the ERC-P-ReXS project (2016-0790). R.A. is supported by the Swedish Research Council 
(VR) under the project ``Evolution of nanoscale charge order in superconducting YBCO nanostructures''. G.S. acknowledges 
support from the Deutsche Forschungsgemeinschaft.

\vspace{0.5truecm}
\noindent
{\bf {Author contributions}}
\par\noindent
S.C., C.D.C, and M.G. conceived the project. G.S. performed the theoretical calculations of the self-energy and
resistivity, with contributions from S.C., C.D.C, and M.G.. R.A., R.F., Y.Y.P, L.B., M.G., and G.G. provided the 
RXS experimental data. M.G., S.C., R.A., L.B., and G.G. performed the fitting of the RXS data. 
The manuscript was written by S.C., C.D.C., M.G., G.S., R.A., and G.G., with contributions and suggestions 
from all coauthors.
~
\vspace{0.5truecm}
\par\noindent
{\bf {Competing interests}}
\par\noindent
The authors declare no competing interests.

\vfill \eject \newpage

\begin{widetext}

{\centerline{\bf{Supplementary Information: Strange metal behaviour from charge density fluctuations in cuprates}}}

\vspace{1cm}
\noindent
{\bf Supplementary Note 1}
\par\noindent
{\bf \small Calculation of the self-energy} 
\vspace{1cm}

We carried out a perturbative calculation of the self-energy corrections of the fermion quasiparticles 
using the Feynman diagram of supplementary figure \ref{fit-selfen}, where the solid line represents a bare quasiparticle, 
and the wavy line may alternatively represent a CDF or a CDW collective excitation.
 
\begin{figure}[htbp]
\includegraphics[angle=0,scale=0.5]{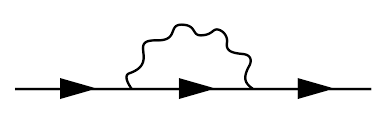}
\caption{Supplementary figure 1. Feynman diagram of the electron self-energy at the lowest perturbative order. The solid lines represent 
the electron propagator, while the wavy line represents either the CDF or the CDW correlator.}
\label{fit-selfen}
\end{figure}

The analytic expression for the (retarded) imaginary part is (see supplementary reference \onlinecite{mazza}) 
\begin{eqnarray}
  \mbox{Im}\,\Sigma({\bf k},\omega)&=& {-}g^2 \int\frac{\mathrm d^2{\bf q}}{(2\pi)^2} \label{eq:self}\\
 &\times&   \frac{(\omega-\varepsilon_{\bf k-q})\lbrack b(\varepsilon_{\bf k-q})+
      f(\varepsilon_{\bf k-q}-\omega)\rbrack}{\lbrack \omega_0+\bar\nu\,\eta_{\bf q}-(\omega-
      \varepsilon_{\bf k-q})^2/\overline\Omega\rbrack^2+(\omega-\varepsilon_{\bf k-q})^2} 
      {\exp(-\eta_{\bf q}/\Lambda)},\nonumber
\end{eqnarray}
where $b(z)=[\mathrm e^{z/k_{\mathrm B} T}-1]^{-1}$ is the Bose function, 
$f(z)=[\mathrm e^{z/k_{\mathrm B} T}+1]^{-1}$ is the 
Fermi function, $g$ is the coupling between electrons and CDFs or CDWs, and 
$(2\pi)^2\eta_q=4-2\cos(q_x-Q^c_x)-2\cos(q_y-Q^c_y)$ contains the information about the CDW/CDF 
vector $\mathbf Q_c$. The function $\eta_q$ is scaled by $1/(2\pi)^2$ because in the fit to RXS the 
wavevector is defined in r.l.u. (see supplementary note 2). For the evaluation of $\Gamma_\Sigma$ we sum over all 
$4$ equivalent wavevectors $(\pm Q_c,0)$ and $(0,\pm Q_c)$, with $Q_c\approx 0.3$\,r.l.u. 
Following supplementary reference \onlinecite{mazza}, we introduce an exponential cutoff which accounts for the suppression 
of the coupling between CDFs/CDWs and quasiparticles away from ${\bf Q}_c$: $\Lambda=0.1$ for CDF scattering 
and $\Lambda=0.5$ for CDW scattering. The electron dispersion $\varepsilon_{\bf k}$ is taken from 
supplementary reference \onlinecite{meevasana}.

In supplementary figure \ref{fit-isotropic} we show that the
scattering due to CDFs stays isotropic even for electron states away from the Fermi surface.

\begin{figure}[htbp]
\includegraphics[angle=0,scale=0.5]{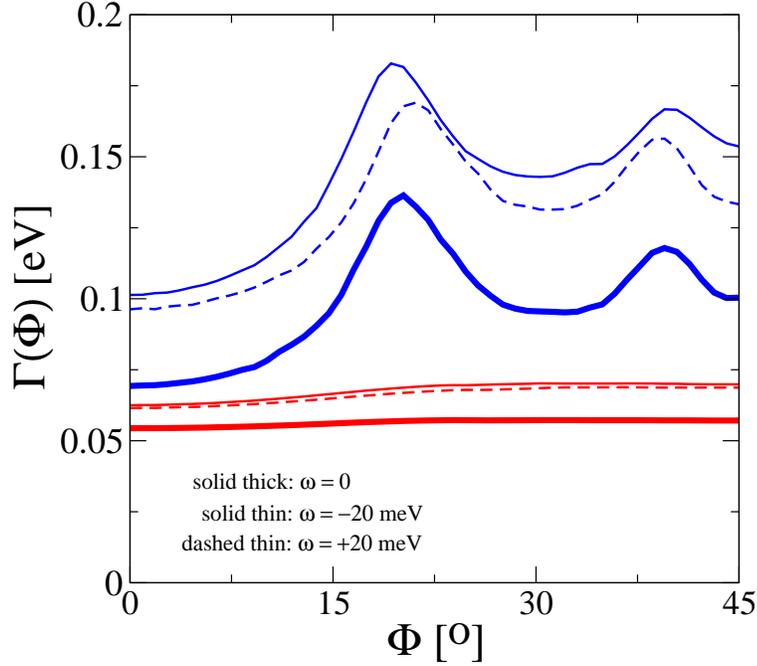}
\caption{Supplementary figure 2. Scattering due to CDWs (blue) and CDFs (red) for electron states at the 
Fermi surface (solid thick lines) and
20\,meV below/above the Fermi surface (solid/dashed thin line). }
\label{fit-isotropic}
\end{figure}

From the real part of the self-energy it is possible to calculate the dimensionless coupling between 
the quasiparticles and the charge excitations. Specifically, using 
$$
\left. \frac{\partial \Sigma'(\omega)}{\partial \omega}\right|_{\omega=0}=\lambda,
$$
with the parameters used to fit the resistivity data, we find $\lambda\approx 0.35-0.5$. 
This value of $\lambda$ nearly doubles at $T=0$, and adding to it the effect of CDWs scattering, 
one obtains a larger value of order $1.5-2.0$ in agreement with values extracted at low temperature from the 
kinks in the electron dispersion measured in photoemission experiments (see supplementary references \onlinecite{mazza,lanzara}).
\par\noindent
{\it --- Temperature dependence}
\par
At $\omega=0$ the self-energy, supplementary equation (\ref{eq:self}), can be rewritten as
\begin{eqnarray*}
\mbox{Im}\,\Sigma({\bf k},\omega=0)&=&{-}g^2\int \mathrm d x \frac{x}{\sinh \frac{x}{k_{\mathrm B} T}} 
f_{\bf k}(x) \\
\mbox{with}\hspace*{0.3cm}  f_{\bf k}(x)&=& \int\frac{\mathrm d^2{\bf q}}{(2\pi)^2} 
\frac{\delta(x-\varepsilon_{\bf q})\,\exp(-\eta_{\bf k-q}/\Lambda)}{\lbrack 
\omega_0+\bar\nu\,\eta_{\bf k-q}-x^2/\overline\Omega\rbrack^2+x^2}\,.
\end{eqnarray*}
For constant parameters $\omega_0$, $\bar{\nu}$, $\overline{\Omega}$ (i.e., when the scattering is dominated by 
CDFs) the temperature dependence of $\mbox{Im}\,\Sigma$ arises from the $x/\sinh\frac{x}{k_B T}$ factor. The 
latter corresponds to a bell-shaped curve, centered at $x=0$ with a half-width at half-maximum of 
$\approx 2 k_{\mathrm B} T$. Since also $f_{\bf k}(x)$ is only finite around $x=0$ (width $w_F$) the temperature 
dependence for $2k_B T < w_F$ is determined by the $x/\sinh\frac{x}{k_{\mathrm B} T}$ factor, yielding 
$\mbox{Im}\,\Sigma \sim \int\! \mathrm dx\,\, x/\sinh \frac{x}{k_{\mathrm B} T}\sim (k_B T)^2$.
On the other hand, for $2k_{\mathrm B} T > w_F$, the integral is cut by the width of $f_{\bf k}(x)$ and 
therefore one can expand $\sinh\frac{x}{k_{\mathrm B} T}$ for $k_{\mathrm B} T\gg x$, which yields a linear 
temperature dependence.

For $\bar{\nu}=0$, it is straightforward to show that the width of $f_{\bf k}(x)$ is determined by 
$\omega_0$, see supplementary figure \ref{fig:fkx}(a). A finite $\bar{\nu}$ introduces a background to 
the function $f_{\bf k}(x)$, due to the contribution of scattering processes where 
${\bf k}-{\bf q} \ne {\bf Q}_c$. This effectively reduces the influence of $\omega_0$ on defining the 
crossover from $T^2$ to linear $T$ behaviour. As can be seen from panel (b), the van Hove singularity (vHs) 
induces an additional feature in $f_{\bf k}(x)$, when the chemical potential is sufficiently close and 
therefore contributes to the weight which is picked up by $x/\sinh \frac{x}{k_{\mathrm B} T}$.

\begin{figure}[h!]
\includegraphics[angle=0,scale=0.6]{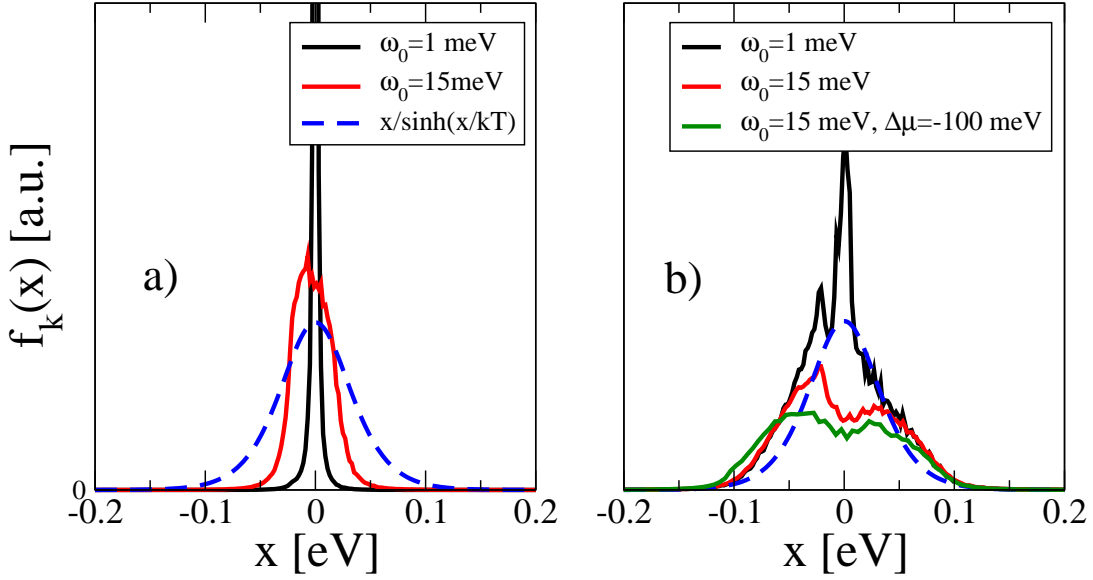}
\caption{Supplementary figure 3. The function $f_{\bf k}(x)$ at the nodal point, together with $x/\sinh\frac{x}{k_{\mathrm B} T}$
at $T=200$\,K. Panel (a) is for $\bar{\nu}=0$, where the width of $f_{\bf k}(x)$
is determined by $\omega_0$. Panel (b) shows results for $\bar{\nu}=1.4$\,eV/(r.l.u.)$^2$. Other parameters: 
$\overline{\Omega}=30$\,meV, $|{\bf Q}_c|=0.3$\,r.l.u. The electron dispersion is from 
supplementary reference \onlinecite{meevasana}, and the chemical potential is $20$\,meV below the vHs, except for the green 
curve in panel (b) where it is $100$\,meV below.}
\label{fig:fkx}
\end{figure}

\par\noindent
{\it --- Frequency dependence}
\par
In case of $\bar{\nu}=0$ (overdamped Holstein model) one can
derive an analytical expression for $\mbox{Im}\,\Sigma(\omega)$
in the limit of zero temperature and a constant density of states $\rho^{2D}$. One obtains
  \begin{equation}
    \mbox{Im}\,\Sigma(\omega)=\left\{\begin{array}{r@{\quad \mbox{for} \quad}l}
    {-}\frac{g^2\rho^{2D}}{\sqrt{\frac{4\omega_0}{\overline{\Omega}}-1}}
  \mbox{atan}\left(\sqrt{\frac{4\omega_0}{\overline{\Omega}}-1}\frac{\omega^2}{2\omega_0^2+
    (1-\frac{2\omega_0}{\overline{\Omega}})\omega^2}\right)\!, & \frac{4\omega_0}{\overline{\Omega}}>1, \\
  {-}\frac{g^2\rho^{2D}}{\sqrt{1-\frac{4\omega_0}{\overline{\Omega}}}}
  \mbox{atanh}\left(\sqrt{1-\frac{4\omega_0}{\overline{\Omega}}}\frac{\omega^2}{2\omega_0^2+
    (1-\frac{2\omega_0}{\overline{\Omega}})\omega^2}\right)\!, & \frac{4\omega_0}{\overline{\Omega}}<1,
  \end{array} \right.
\end{equation}
which, for the fitted CDF parameters ($\omega_0/\overline{\Omega}=\frac{1}{2}$, see supplementary note 2), reduces 
to
\begin{equation}
  \label{eq:self-1}
    \mbox{Im}\,\Sigma(\omega)={-}g^2\rho^{2D}
  \mbox{atan}\left(\frac{\omega^2}{2\omega_0^2}\right)\!.
\end{equation}
The function in supplementary equation (\ref{eq:self-1}) displays a quadratic behaviour up to $\omega \approx \omega_0$, before it enters 
into an extended linear regime. Repeating the analysis with a lattice 2D DOS yields an additional hump in 
$\mbox{Im}\,\Sigma(\omega)$, due to the vHs. The energy $\omega$ of this hump is ruled by the distance of the
chemical potential from the vHs, but also affected by the excitation frequency $\omega_0$. As argued before, a 
finite $\bar{\nu}$ introduces scattering processes away from $Q_c$, which effectively enhance $\omega_0$ and therefore 
move the hump to higher energies [in Fig.\,2(c) of the main manuscript the hump is visible at $\approx 0.15$\,eV]. For 
small frequencies and in the strongly damped limit $\overline{\Omega}\to \infty$, it has been shown in 
supplementary reference \onlinecite{caprara99} that Fermi liquid behaviour $\sim \omega^2$ persists up to
$\omega_{FL}\approx \omega_0+\frac{1}{2}\bar{\nu}a^2 (\varepsilon_{k-Q_c}/v_{k-Q_c})^2$,
where $a$ is the lattice constant and $v_k$ denotes the Fermi velocity. For the overdoped sample this estimate yields 
$\omega_{FL}\approx 30$\,meV, in good agreement with what is seen in Fig.\,2(c) of the main manuscript.

\newpage
\noindent
{\bf Supplementary Note 2}
\par\noindent
{\bf \small Extracting the CDF and CDW dynamics from RXS spectra} 
\vspace{1cm}

As mentioned in the Methods section of the main manuscript, the CDW or CDF contribution to the low-energy RXS spectra is 
\beq
I(\qvec,\omega)=A\,\mathrm{Im}\,D(\qvec,\omega)\,b(\omega),
\label{teo-HR-SI}
\eeq
where $D$ is the fluctuation propagator [see below, supplementary equation (\ref{fluctuator})], 
$b(\omega)\equiv[\mathrm e^{\omega/k_{\mathrm B} T}-1]^{-1}$ is the Bose distribution, 
$k_{\mathrm B}$ is the Boltzmann constant, and $A$ is a constant effectively representing 
the intricate photon-conduction electron scattering processes (see supplementary references 
\onlinecite{reviewRIXS1,reviewRIXS2}). This contribution 
to RXS spectra corresponds to the Feynman diagram of supplementary figure \ref{Fig1-MFL-SI}(a). 
The shaded rectangles represent the coupling between the incoming and outgoing photons (dashed blue lines) 
with the conduction electrons (solid blue lines). These rectangles schematise the complicated processes 
underlying the RXS scattering (see supplementary references 
\onlinecite{reviewRIXS1,reviewRIXS2}): the incoming photon creates a core hole and an excited 
electron in some ${\bf k}$ conduction band state. Another conduction electron in a ${\bf k-q}$ state fills 
the core hole emitting the outgoing photon. The system is then left with a conduction electron-hole pair with 
momenta ${\bf k}$ and ${\bf k-q}$. Of course, the shaded rectangles also schematise the intermediate 
interaction processes between the core hole and the surrounding electrons, the other electron-electron 
interactions and so on. At low energy the particle-hole pair is usually in the conduction band and it may 
decay in the collective excitations of our interest, which are represented by the wavy line in supplementary figure 
\ref{Fig1-MFL-SI}(a). 
In the low-energy range ($|E|<0.1$\,eV), since we focus on the momentum and energy dependence of the low-energy 
collective modes, all the above high-energy intermediate complicated processes may be represented by 
an effective constant $A$ in the RXS response function.

\begin{figure}[htbp]
\includegraphics[angle=0,scale=0.4]{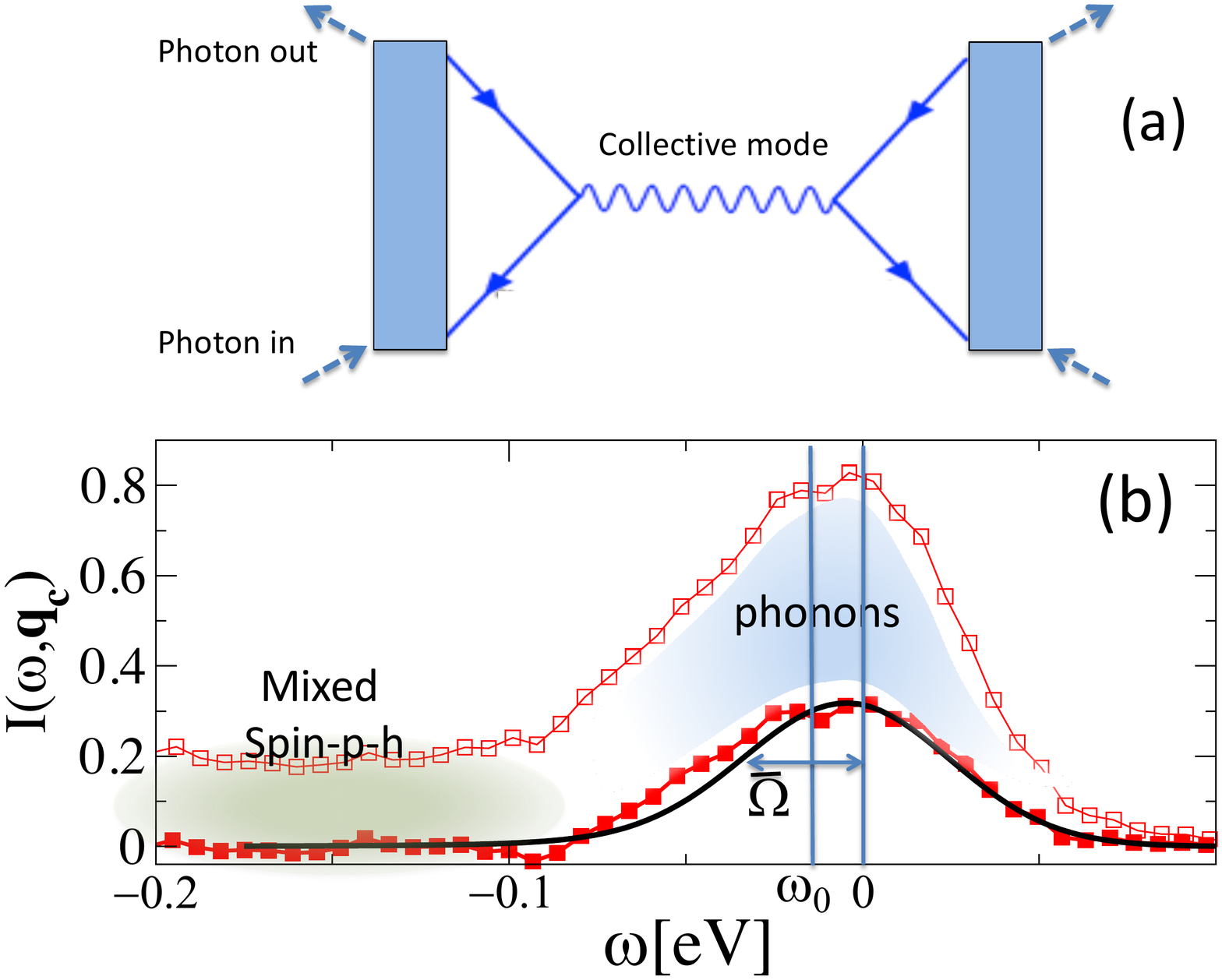}
\caption{Supplementary figure 4. Feynman diagram schematising the contribution of a charge collective mode (wavy line) to the 
RXS spectra. The shaded rectangles schematically represent the coupling between the incoming and outgoing 
photons (dashed blue lines) with the conduction electron-hole pairs (solid blue lines). These shaded 
rectangles encode the complicated processes with core-hole intermediate states (see text).
(b) Example of an high-resolution RXS spectrum of an optimally doped  ($T_{\mathrm c}=90$\,K) 
Nd$_{1+x}$Ba$_{2-x}$Cu$_3$O$_{7-\delta}$ sample at $T=250$\,K and at the critical wavevector for CDFs 
$\qvec=\Qvec_c\approx (0.3,0)$\,r.l.u. (empty squares). The contribution of phonon excitations and of the 
spin and particle-hole excitations are shaded in light blue and green, respectively. The pure CDF 
spectrum obtained subtracting these contributions is reported with filled squares.}
\label{Fig1-MFL-SI}
\end{figure}
    
The latter is connected, via fluctuation-dissipation theorem, to the dissipative (i.e., imaginary) part 
of the dynamical density response function $D(\qvec,\omega)$, which we take in the Gaussian approximation 
of the linear response theory as
\beq
D(\qvec,\omega)\equiv \sum_{n=1}^{4} 
\left[\omega_0+\bar{\nu}\eta_n(\qvec)-\mathrm i\omega -
\frac{\omega^2}{\overline\Omega} \right]^{-1},
\label{fluctuator}
\eeq
describing either CDWs or CDFs [see Eq.\,(3) of the main text].
The function $\eta_q$ is scaled by $1/(2\pi)^2$ in order to make our results (in particular the value for 
$\bar{\nu}$) compatible with supplementary reference \onlinecite{arpaia-2018}, where the fit to RXS is performed with the continuum 
version of supplementary equation (\ref{fluctuator}), i.e., $\eta_n({\bf q})\to (q_x-Q^c_x)^2+(q_y-Q^c_y)^2$ and with 
the wavevector defined in r.l.u.
The sum in supplementary equation (\ref{fluctuator}) runs over the four equivalent peaks along the H- and K-direction
corresponding to  the four equivalent CDW/CDF vectors $(\pm Q_c,0)$, $(0,\pm Q_c)$. In supplementary equation (\ref{fluctuator}) 
we have implemented a lattice periodic function $(2\pi)^2\eta_n({\bf q})=4-2\cos(q_x-Q^c_x)-2\cos(q_y-Q^c_y)$
so that peaks occur in all Brillouin zones. The fact that $\eta(\qvec)$ is periodic and does not grow much away 
from the critical wavevectors implies that $D$ has  sizeable weight even away from the $\Qvec_c$'s and a substantial 
background is overall present in the whole Brillouin zone. To extract more specific information from the observed 
narrow and broad peaks (henceforth, NP and BP respectively), the spectra along the (H,H) direction have been subtracted from those 
obtained with scans along the (H,0) direction. This of course eliminates the rather uniform background due to the 
periodic form of $\eta(\qvec)$ and the difference spectra are more properly fitted with the simpler and more
transparent quadratic continuum dispersion of the modes around $\Qvec_c$.

At $\qvec=\Qvec_c$ high-resolution RXS spectra have the form reported in supplementary figure \ref{Fig1-MFL-SI}(b). 
Once the spin and phonon contributions are subtracted, valuable information can be extracted to determine 
the dynamics of the CDFs and CDWs. Specifically, using supplementary equation (\ref{teo-HR-SI}), one can fit the high-resolution 
spectra at high temperature (where CDWs are not present) to find the dynamical scale $\omega_0$ of CDFs. Then, 
this information can be used to fit the quasi-elastic peaks and extract the relative weight (intensity) of the 
narrow and broad contributions at all temperatures. Once this information is obtained, the relative weight of 
the CDF and CDW contribution at all temperatures is known, and one can go back to high-resolution spectra at 
lower $T$. This bootstrap approach is needed at temperatures $T\lesssim T^*$ where both 
CDFs and CDWs are present. However, we emphasise that our main goal is to identify the scattering mechanism 
responsible for the strange-metal behavior occurring {\it above} $T^*$, where CDFs only are present and the 
involved attempt of separating CDW and CDF contributions to the RXS spectra is not in order.
For completeness, though, we hereafter describe the complete procedure in detail.

The peak in the quasi-elastic RXS spectra has a composite character and, once the background 
measured along the $(1,1)$ direction is subtracted (see, e.g., Fig.\,2 A-D in supplementary reference \onlinecite{arpaia-2018}), 
it may be decomposed into two approximately Lorentzian contributions. The narrow peak, a strongly temperature 
dependent peak, is due to the well-known nearly critical CDWs arising below $T_{\mathrm{CDW}}\approx 170-200$\,K 
(for the sample at optimal doping),  while a broad peak is also present due to the CDFs. 
The identification of this broad peak is the main outcome of the RXS experiments reported in 
supplementary reference \onlinecite{arpaia-2018}. We fitted the experimental data (blue points in supplementary figure \ref{Fig2-MFL-SI}) 
with Eq.\,(4) in Methods. 

\begin{figure}[htbp]
\includegraphics[angle=0,scale=0.5]{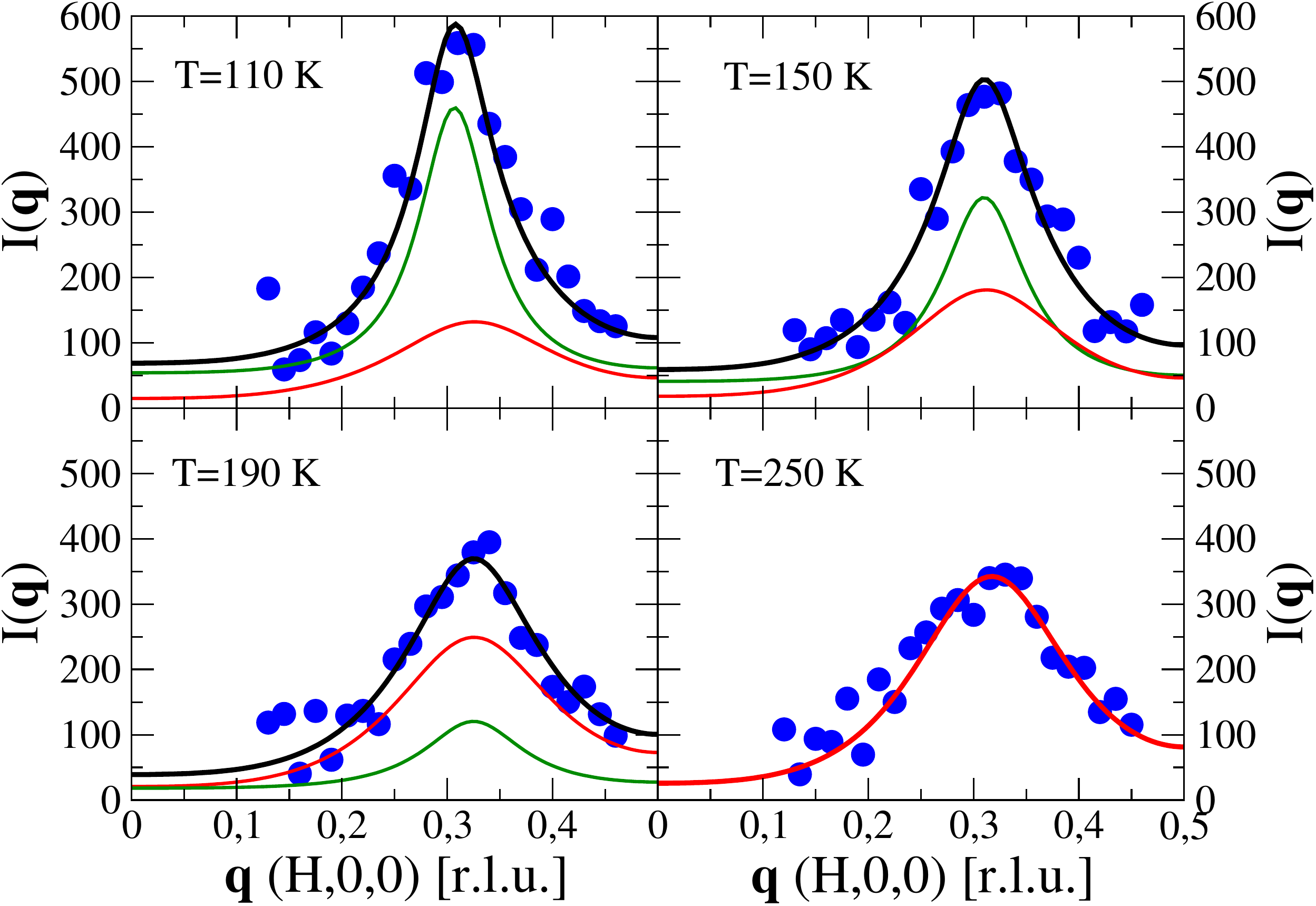}
\caption{Supplementary figure 5. Quasi-elastic RXS spectra along the (H,0) direction of an optimally doped 
($T_{\mathrm c}=90$\,K) Nd$_{1+x}$Ba$_{2-x}$Cu$_3$O$_{7-\delta}$ sample (blue dots), at four different 
temperatures above $T_{\mathrm c}$. The subtraction of the linear background measured along the (1,1) 
direction has been performed. The fitting curve (black solid line is the sum of two contributions: a narrow 
peak (attributed to nearly critical CDWs, green solid lines) and a broad peak due to CDFs (red solid line).}
\label{Fig2-MFL-SI}
\end{figure}

From the fits one can extract for each of the two, NP and BP, components, the overall intensity parameter 
$A$ and the ratio $\omega_0/\bar\nu$. Since only this ratio determines the width of the quasi-elastic spectra, 
$\sqrt{\omega_0/\bar\nu}=\xi^{-1}$, we make use of a separate measure of high-resolution spectra to disentangle 
$\omega_0$ and $\bar\nu$. Therefore for the optimally doped sample with $T_{\mathrm c}=90$\,K we used the 
high-resolution information on $\omega_0$ for the broad peak at $T=250$\,K to extract 
{{$\bar\nu_{\mathrm{BP}}\approx 1400$\,meV/(r.l.u.)$^2$}} at these temperatures. The same procedure cannot be adopted 
for the narrow CDW peaks, which always appear on top of the broad CDF contribution. Nevertheless, to obtain a 
rough estimate, we investigated the high-resolution spectra at low temperature, where the maximum intensity 
should mostly involve the narrow peak to extract the characteristic energy of the quasi-critical CDWs obtaining, 
as expected, much lower values $\omega_0^{\mathrm{NP}}\approx 1-3$\,meV (although these low values are less reliable, due 
to the relatively low resolution of the frequency-dependent spectra).

\begin{figure}[htbp]
\includegraphics[angle=0,scale=0.3]{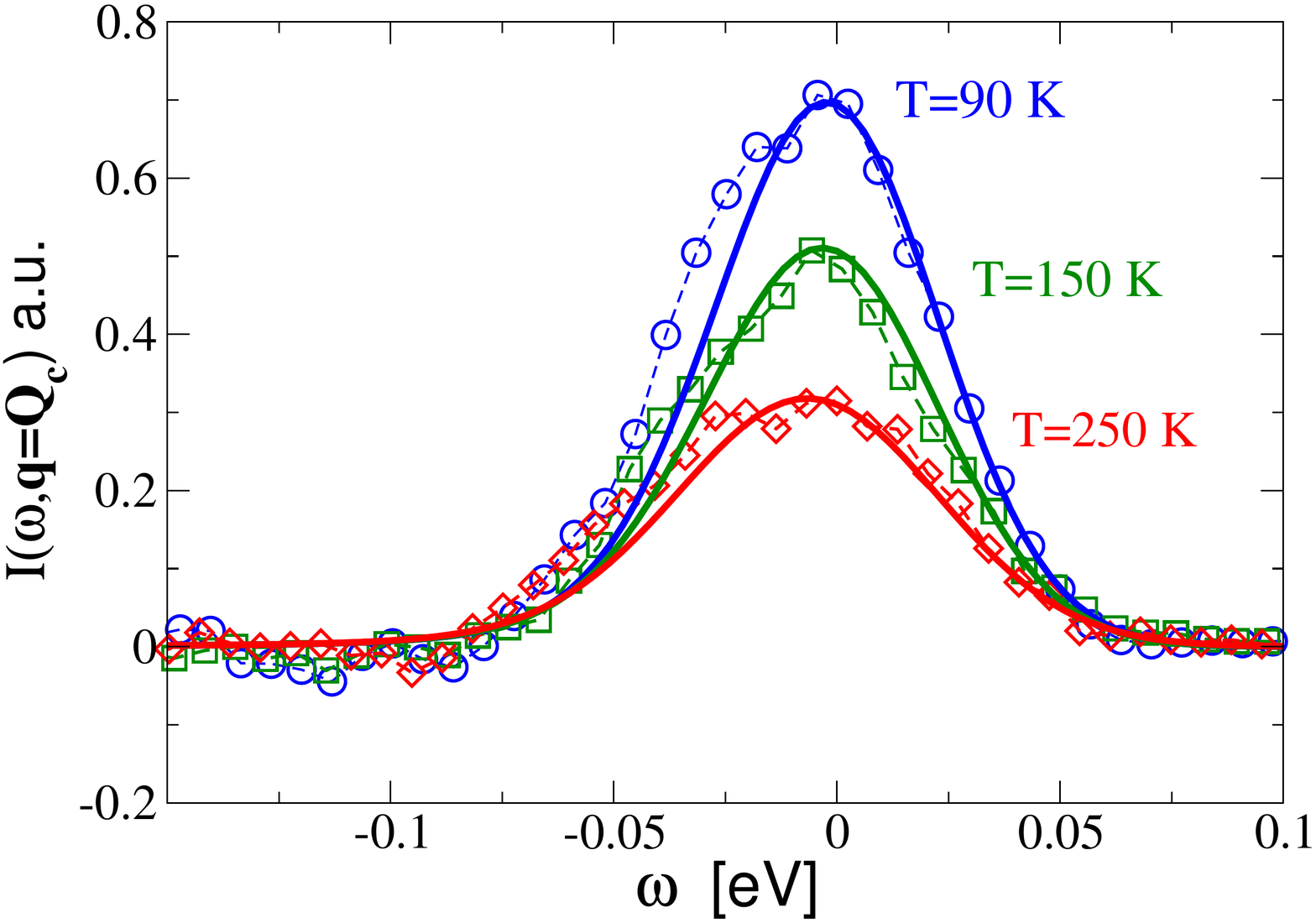}
\caption{Supplementary figure 6. High-resolution RXS spectra of an optimally doped ($T_{\mathrm c}=90$\,K)  
Nd$_{1+x}$Ba$_{2-x}$Cu$_3$O$_{7-\delta}$ sample. At all temperatures the fitting parameters are:
$\omega^{\mathrm{BP}}_0 =15$ meV,  $\bar{\nu}_{\mathrm{BP}}=1400 $ meV,  $\bar{\nu}_{\mathrm{NP}}=800 $ meV,
$\overline{\Omega}_{\mathrm{NP},\mathrm{BP}} =30$ meV. Only the CDW characteristic energy varies 
with $T$:  $\omega_0^{\mathrm{NP}}(90K) =0.9$ meV,  $\omega_0^{\mathrm{NP}}(150K) =1.9$ meV. At $T=250$ K no
CDW peak is present. The same relative weight of the NP and BP as determined from
low-resolution spectra has been used here.
Only a factor common to all theoretical curves for the overall intensity has been adjusted. The 
horizontal axis of the experimental data has been slightly shifted (but within the estimated $\pm 4$\,meV 
energy error bar) to match the right part of the experimental and theoretical spectra in the region where 
the Bose distribution factor cuts the positive excitation energies: $-2$\,meV for the $T=90$\,K data, 
$+4$\,meV for the $T=150$\,K data, and  $-2$\,meV for the $T=250$\,K data.}
\label{hr-spectra}
\end{figure}

These estimates allow to extract values of $\bar\nu_{\mathrm{NP}}\approx 800$\,meV/(r.l.u.)$^2$ for the 
CDWs, comparable to those of the CDFs. This suggests that common electron degrees of freedom (e.g., the 
fermion quasiparticles in the approach of supplementary references \onlinecite{CDG-1995,reviewQCP1,CDSG-2017}) 
underlie both kinds of charge density excitations. 
This fitting procedure allows to identify the relative intensity $A_{\mathrm{NP}},A_{\mathrm{BP}}$ of the narrow and broad peaks. 
To reduce the fitting parameters to a minimum, although 
sub-leading temperature dependencies of the high-energy parameters $\bar\nu$ and $\overline\Omega$ over a 
broad temperature range can be expected, we kept those parameters constant. We also assumed a constant 
$\omega_0$ for the CDFs, to highlight the non-critical nature of these fluctuations.  
Of course, introducing a mild (i.e., non critical) temperature dependence of the parameters like 
$\bar\nu, \bar\Omega,\omega_0^{\mathrm{BP}}$ can only improve the fits and moderately alter the relative weights 
of the NP and BP components.

Once the fitting of the quasi-elastic spectra was carried out, we analysed the high-resolution RXS spectra 
at lower temperatures. At this stage, but for a common overall factor (the overall intensity of 
high-resolution spectra being unrelated to the intensity of the low-resolution spectra, due to the 
different time and conditions of the corresponding measures) we have no more free parameters to use 
because the characteristic energies and the relative weight of the CDWs and CDFs were determined. 
Subtracting the phonon and spin/particle-hole excitations [as determined from the high-resolution spectra 
at $\qvec \parallel (1,1)$], one obtains spectra where the contribution of CDFs and CDWs is only (or 
predominantly) present. Then we obtained spectra and the fits of supplementary figure \ref{hr-spectra}.

These spectra are taken at three representative temperatures: at high temperature
($T=250$\,K), where only CDFs are present, at intermediate temperature ($T=150$\,K), where CDFs and 
CDWs coexist, and at low temperature ($T=T_{\mathrm c}=90$\,K), where the CDWs are more pronounced than CDFs 
(see also supplementary figure \ref{Fig2-MFL-SI}).
  
We stress once more that the above analysis faces the difficult issue of identifying and separating 
the contribution of the CDFs and CDWs to produce the NP and BP observed in the RXS spectra. While this issue 
will be addressed in a separate work, where the effects of combined scattering of both CDFs and CDWs will be studied, 
for the present purposes of identifying the scatterer responsible for the strange metal behavior,  we only consider 
the $T>T^*$ region of the phase diagram. Here, only CDFs are present and the RXS spectra only have the BP component 
so that the separation of the NP and BP component of the spectra is no longer in order. In this way we will achieve 
our goal, which is to show that the CDFs are able to account for the strange-metal behavior.
\newpage
\noindent
{\bf Supplementary Note 3}
\par\noindent
{\bf \small Calculation of the resistivity} 
\vspace{1cm}

The results for the in-plane resistivity presented in the paper are obtained within a
Boltzmann-equation approach, following the derivation of supplementary reference \onlinecite{hussey}. 
We obtain
\begin{equation}
\frac{1}{\rho}=\frac{e^2}{\pi^3\hbar}\frac{2\pi}{d}\int\!\mathrm d\phi\,
\frac{k_F(\phi) v_F(\phi) \cos^2(\phi-\gamma)}{\Gamma(\phi)\cos(\gamma)}\, ,\label{eq:rhoxx}
\end{equation}
where $k_F(\phi)$, $v_F(\phi)$, and $\Gamma(\phi)$ 
denote the angular dependence of the Fermi momentum, Fermi velocity, 
and scattering rate along the Fermi surface [see Fig.\,2(a) of the main text], and 
\begin{displaymath}
\gamma=\mbox{atan}\left(\frac{1}{k_F}\frac{\partial k_F}{\partial\phi}\right)\,.
\end{displaymath}
The scattering rate $\Gamma(\phi)\equiv\Gamma_0 + \Gamma_{\Sigma}(\phi)$ includes 
an elastic scattering rate $\Gamma_0$, and the scattering rate due to CDWs or CDFs,
$\Gamma_{\Sigma}(\phi)\equiv{-}\mathrm{Im}\,\Sigma(k_F(\phi),\omega=0)$, where
$\Sigma(\mathbf k,\omega)$ is the retarded electron self-energy (see supplementary note 1).

The electron dispersion $\varepsilon_{\bf k}$ includes nearest-, next-nearest- and next-next-nearest-neighbor 
hopping terms generic for cuprates (see supplementary reference 
\onlinecite{meevasana}).
The in-plane lattice constant for YBCO is taken as $a=3.85$\,\AA \, and the $c$-axis lattice constant 
is $d=11.7$\,\AA. The bilayer structure of YBCO is effectively taken into account by multiplying 
supplementary equation (\ref{eq:rhoxx}) by a further factor of 2.

\newpage
\noindent
{\bf Supplementary Note 4}
\par\noindent
{\bf \small Linear response theory} 
\vspace{1cm}

Alternatively to supplementary equation (\ref{eq:rhoxx}), one can also adopt a Kubo approach for the evaluation of 
$\rho$. Following supplementary reference \onlinecite{allen15}, the conductivity (e.g., along the $x$-direction) can be 
obtained from
\begin{equation}\label{eq:allen}
  \sigma_{xx}(\omega+\mathrm i\eta)=\frac{\mathrm i e^2}{\omega}\int_{-\infty}^{+\infty} \mathrm d\nu
  \left\lbrack f(\nu)-f(\nu+\omega)\right\rbrack
  \sum_k \frac{v_{kx}^2\delta(\varepsilon_k-\mu)}{\omega-\Sigma(k,\nu+\omega+\mathrm i\eta)
  +\Sigma^* (k,\nu+\mathrm i\eta)},
\end{equation}
where $f(\nu)\equiv[\mathrm e^{(\nu-\mu)/k_{\mathrm B} T}+1]^{-1}$ is the Fermi distribution, $\mu$ is the
chemical potential, and $v_{kx}$ denotes the Fermi velocity along the $x$-direction.

\begin{figure}[htbp]
\includegraphics[angle=0,scale=0.4]{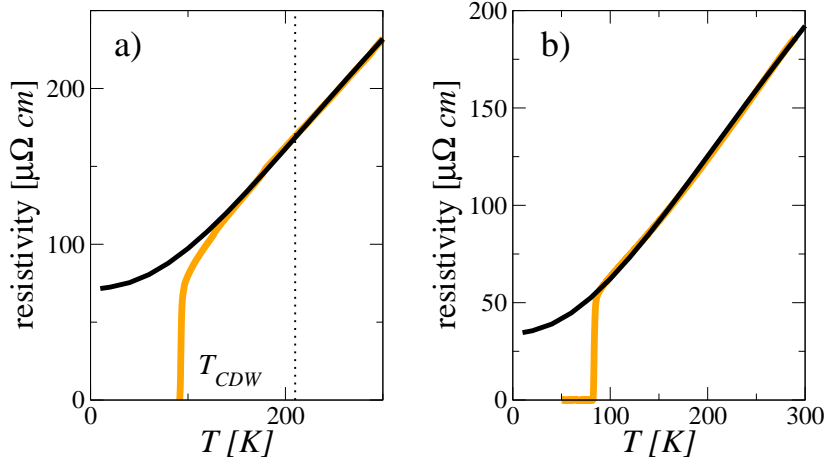}
\caption{Supplementary figure 7. (a) Experimental resistivity for an optimally doped ($T_{\mathrm c}=90$\,K) NBCO sample (yellow 
thick curve compared to the theoretical result obtained from supplementary equation (\ref{eq:allen}), within a Kubo approach. 
The scattering due to quenched impurities is included via $\eta \to \eta+\Gamma_0$, where $\Gamma_0$ is the 
elastic scattering rate. Coupling to quasiparticles $g=0.18$\,eV, $\Gamma_0=33$\,meV. (b) Same as (a), for 
an overdoped YBCO sample ($T_{\mathrm c}=83$\,K), with $g=0.168$\,eV, $\Gamma_0=16$\,meV.
}
\label{kubo}
\end{figure}

Supplementary figure \ref{kubo} shows fits for the resistivity which can be directly compared with Fig.\,3 of the main 
manuscript that have been obtained from supplementary equation (\ref{eq:rhoxx}). For the overdoped sample 
[Fig.\,3(b) of the main manuscript and supplementary figure \ref{kubo}(b)] the Kubo
approach yields a slightly better agreement close to $T_{\mathrm c}$ (albeit, also neglecting vertex 
corrections, and with the momentum dependence perpendicular to the Fermi surface being still approximative). 
Nevertheless, the fact that both Kubo and Boltzmann approach yield rather good agreement with experiment 
supports the evidence that CDFs can account for the linear-in-$T$ behaviour of the electron self-energy.

\end{widetext}
\vfill \eject

\newpage
\noindent
{\bf Supplementary references}

\end{document}